\documentclass[pre,twocolumn]{revtex4-1}
\usepackage{amsmath,amssymb,bm,natbib}
\usepackage{hyperref}

\newcommand{\Tr}{{\rm Tr}}

\newcommand{\be}{\begin{equation}}
\newcommand{\ee}{\end{equation}}

\begin{document}

\title{Semiclassical approach to $S$ matrix energy correlations and time delay in chaotic systems}
\author{Marcel Novaes}
\affiliation{Instituto de F\'isica, Universidade Federal de Uberl\^andia, Uberl\^andia, MG, 38408-100, Brazil}
\date{\today}

\begin{abstract}

The $M$-dimensional scattering matrix $S(E)$ which connects incoming to outgoing waves in a chaotic systyem is always unitary, but shows complicated dependence on the energy. This is partly encoded in correlators constructed from traces of powers of $S(E+\epsilon)S^\dagger(E-\epsilon)$, averaged over $E$, and by the statistical properties of the time delay operator, $Q(E)=-i\hbar S^\dagger dS/dE$. Using a semiclassical approach for systems with broken time reversal symmetry, we derive two kind of expressions for the energy correlators: one as a power series in $1/M$ whose coefficients are rational functions of $\epsilon$, and another as a power series in $\epsilon$ whose coefficients are rational functions of $M$. From the latter we extract an explicit formula for $\Tr(Q^n)$ which is valid for all $n$ and is in agreement with random matrix theory predictions. 

\end{abstract}

\maketitle

\section{Introduction}

Scattering of waves of energy $E$ can be described by the $S(E)$ matrix, which
connects incoming to outgoing amplitudes. We consider a finite region with chaotic classical dynamics, characterized by a single time scale $\tau_D$, the dwell time, the average amount of time spent inside the region by a classical particle injected at random. This chaotic region is connected to the outside world by means of $M$ channels, so that $S$ is $M$-dimensional and always unitary as a consequence of the energy conservation. 

If time-reversal symmetry is broken, one statistical approach, random matrix theory (RMT), assumes $S(E)$ to be uniformly distributed in the unitary group \cite{uzy1,uzy2}, according to the invariant Haar measure, for every $E$. To understand the correlations between $S$ matrices at different energies has always been a challenge. One way to quantify this is to compute $S$ at one energy and $S^\dagger$ at another, and take the trace of their product,
$ \Tr\left[S\left(E+\frac{\epsilon\hbar}{2\tau_D}\right)S^\dagger\left(E-\frac{\epsilon\hbar}{2\tau_D}\right)\right]$. This will be equal to $M$ for $\epsilon=0$, but in general a widely fluctuating function of $E$. Averaging within a local energy window produces a well behaved funcion of $\epsilon$. Such energy correlations have traditionally been studied by modelling the Hamiltonian of the system as a random hermitian matrix coupled to scattering channels \cite{verb,dima,yan,dietz1,dietz2,dietz3,dietz4}. 

A more detailed characterization of energy correlations is the calculation of 
\be\label{corr} C_n(M,\epsilon)=\left\langle\Tr\left[S\left(E+\epsilon'\right)S^\dagger\left(E-\epsilon'\right)\right]^n\right\rangle\ee
for integer $n$, where $\epsilon'=\frac{\epsilon\hbar}{2\tau_D}$. The above quantity is expected to be universal, i.e. independent of the system's details as long as it is chaotic. Besides $M$ and $\epsilon$, it should depend only on whether time-reversal symmetry is present or not. In this work we focus our attention on systems where this symmetry is broken.

Related to energy dependence of the $S$ matrix is the time delay matrix \cite{time1,time2,time3,gopar} \be Q(E)=-i\hbar S^\dagger \frac{dS}{dE}.\ee Its real eigenvalues $\{\tau_1,...,\tau_M\}$ are commonly referred to as proper time delays and provide the lifetimes of metastable states. Its normalized trace $\tau_W=\frac{1}{M}\Tr(Q)$ is known as the Wigner time delay, which provides a measure of the density of states of the open system. Its average value equals the classical dwell time, $\langle \tau_W\rangle=\tau_D$. More detailed information is encoded in higher spectral moments such as
\be\label{Qn} Q_n=\langle\Tr(Q^n)\rangle.\ee

The statistical properties of time delay have been much studied. Within RMT, perhaps the main point of departure is the distribution of the inverse matrix $Q^{-1}$, which is known to conform to the Laguerre ensemble \cite{frahm1,frahm2}. This lead to the calculation of the distribution function of $\tau_W$ in different regimes and to expressions for the above spectral moments \cite{majumdar,simm1,simm2,simm3,garcia,novaesrmt,grabsch} (see the review \cite{texier}). 

In this work we do not rely on random matrices, but instead employ a semiclassical approach, in which the elements of $S$ are approximated, in the short-wavelength regime, as infinite sums over scattering rays \cite{miller,raul}. It has been very successful in treating transport properties at fixed energy \cite{essen3,essen5,sieber1,sieber2,greg1,greg2,greg3}. It was adapted by Kuipers and Sieber in order to take into account the variable $\epsilon$ and handle correlators like (\ref{corr}). It has grown into an independent line of attack to this kind of problems \cite{berko1,berko2,andreev1,andreev2,kuipers,kuipersrichter,novaessemi}.

We follow recent advances in the semiclassical theory and formulate correlation functions in terms of auxiliary matrix integrals \cite{matrix,novaessemi,trs,novaessemi2}. These integrals are then computed using Schur polynomials. This leads to two explicit formulas for $C_n(M,\epsilon)$: one as a power series in $1/M$ whose coefficients are rational functions of $\epsilon$, and another as a power series in $\epsilon$ whose coefficients are rational functions of $M$. From the latter we extract an explicit formula for $\Tr(Q^n)$ which is valid for arbitrary values of $n$ and $M$ and which is in agreement with random matrix theory predictions. 

In Section 2 we present the semiclassical matrix integral which is the crux of the theory. In Sections 3 and 4 we use it to compute $C_n(M,\epsilon)$ in two different ways. In Section 5 we make the connection with $Q_n$. We conclude in Section 6.

\section{Semiclassical matrix integrals}

The semiclassical approximation to quantum scattering has been extensively discussed in previous works \cite{essen3,essen5,greg1,novaessemi}. When correlations among scattering trajectories are taken into account, and the required integrations over phase space have been performed, the theory has a diagrammatic formulation which is a perturbative theory in the parameter $M^{-1}$. Kuipers and Sieber obtained the diagrammatic rules governing this theory \cite{KS1,KS2} when applied to (\ref{corr}). The contribution of any given diagram factorizes into the contributions of individual vertices and edges: a vertex of valence $2q$ gives rise to $-M(1-iq\epsilon)$; channels of any valence give rise to $M$; each edge gives rise to $[M(1-i\epsilon)]^{-1}$. 

Recently, the semiclassical approach has been developed in terms of appropriate matrix integrals \cite{matrix,novaessemi,trs,novaessemi2} into which the diagrammatic rules are built by design. For systems with broken time-reversal symmetry, which are our focus, the result is that
\be C_n=\lim_{N\to 0}\int e^{-\sum_{q=1}^\infty \frac{M}{q}(1-iq\epsilon)\Tr(ZZ^\dagger)^q}\Tr[ZPZ^\dagger P]^n\frac{dZ}{\mathcal{Z}},\ee
where $Z$ is an $N$-dimensional complex matrix, $P$ is an orthogonal projector from $\mathbb{R}^N$ to $\mathbb{R}^M$ and \be\mathcal{Z}=\int e^{-M(1-i\epsilon)\Tr(ZZ^\dagger)}dZ\ee is a normalization.

The way this matrix model works is that the factor $e^{-M(1-i\epsilon)\Tr(ZZ^\dagger)}$ is kept as a Gaussian measure while the rest of the exponential is Taylor expanded. Each trace then becomes a vertex in a diagram, along with the correct factor $-M(1-iq\epsilon)$. Then the integration is performed by invoking Wick's rule, and edges are produced along with the correct factor $[M(1-i\epsilon)]^{-1}$. The term $\Tr[ZPZ^\dagger P]^n$ mimicks the correlator we want to compute. Finally, the limit $N\to 0$ is necessary to remove spurious contributions coming from unwanted periodic orbits \cite{matrix}.

The traditional singular value decomposition $Z=UDV^\dagger$, where $U$ and $V$ are unitary matrices, leads to $\mathcal{Z}=\mathcal{G}\int  e^{-M(1-i\epsilon)\Tr(X)}|\Delta(X)|^2dX$, where the Vandermonde
\be \Delta(X)=\prod_{1\le i<j\le N}(x_j-x_i)\ee
is the jacobian of the change of variables and $\mathcal{G}=\int dUdV$ is the result of a double integration over the unitary group. This integral gives
\be \mathcal{Z}=\mathcal{G}(M(1-i\epsilon))^{-N^2}\prod_{j=1}^Nj!(j-1)!.\ee

Let $\chi_\lambda(\mu)$ be the characters of the irreducible representations of the permutation group $S_n$ (these representations are labelled by integer partitions, denoted by $\lambda\vdash n$ or $|\lambda|=n$). They are useful in expressing the trace in terms of Schur polynomials,
\be \Tr(A^n)=\sum_{\lambda\vdash n}\chi_\lambda(n)s_\lambda(A).\ee
Using this fact and the identity
\be \frac{1}{\int dU}\int dU s_\lambda(UAU^\dagger B)=\frac{s_\lambda(A)s_\lambda(B)}{s_\lambda(1^N)}\ee
we get 
 \begin{multline}C_n(M,\epsilon)=\lim_{N\to 0}\sum_{\lambda\vdash n}\left(\frac{s_\lambda(1^M)}{s_\lambda(1^N)}\right)^2\chi_\lambda(n)\\\times\frac{\mathcal{G}}{\mathcal{Z}}\int e^{-M\sum_{q=1}^\infty \frac{1}{q}(1-iq\epsilon)\Tr(X^q)}s_\lambda(X)dX.\end{multline}

The value of the Schur polynomial at an identity matrix is
\be s_\lambda(1^N)=\frac{d_\lambda}{n!}[N]^\lambda,\ee
where $d_\lambda$ is the dimension of the associated irreducible representation, given by
\be \chi_\lambda(1^n)=n!\prod_{i=1}^{\ell(\lambda)}\frac{1}{(\lambda_i-i+\ell)!}\prod_{j=i+1}^{\ell(\lambda)}(\lambda_j-j-\lambda+i),\ee
and $[N]^\lambda$ is a monic polynomial in $N$,
\be [N]^\lambda=\prod_{j=1}^{\ell(\lambda)}\frac{(N+\lambda_j-j)!}{(N-j)!},\ee
which is a generalization of the rising factorial. For future reference, let us also define a corresponding generalization of the falling factorial,
\be [N]_\lambda=\prod_{j=1}^{\ell(\lambda)}\frac{(N-\lambda_j+j)!}{(N+j)!}\ee

Therefore,
\be C_n(M,\epsilon)=\lim_{N\to 0}\sum_{\lambda\vdash n}\left(\frac{[M]^\lambda}{[N]^\lambda}\right)^2\chi_\lambda(n)\mathcal{I}_\lambda,\ee
with 
\be \label{I}\mathcal{I}_\lambda=\frac{\mathcal{G}}{\mathcal{Z}}\int e^{-M\sum_{q=1}^\infty \frac{1}{q}(1-iq\epsilon)\Tr(X)^q}s_\lambda(X)dX.\ee

It is known that $\chi_\lambda(n)$ is different from zero only if $\lambda=(n-k,1^k)$, a so-called hook partition. In that case, $\chi_\lambda(n)=(-1)^k$ and $d_\lambda=\binom{n-1}{k}$. We also define the quantity 
\be t_\lambda=(n-k-1)!k!.\ee
We denote by $H_n$ the set of all hook partitions of $n$. For example, $H_4=\{(4),(3,1),(2,1,1),(1^4)\}.$

\section{Correlator as power series in $1/M$}

Let $b_\beta$ be the size of the conjugacy class of the permutation group containing permutations of cycle type $\beta$ and let us define the function
\be g_\beta(\epsilon)=\prod_{q\in\beta}(1-iq\epsilon).\ee
Then we can expand  $e^{-\sum_{q>1}\frac{M}{q}(1-iq\epsilon)\Tr(X^q)}$ as
\be\label{exp} \sum_{m} \sum_{\beta\vdash m}\frac{1}{m!}b_\beta(-M)^{\ell(\beta)}g_\beta(\epsilon)p_\beta(X),\ee where 
\be p_\beta(X)=\prod_{j=1}^{\ell(\beta)}\sum_{i=1}^Nx_i^{\beta_j}\ee
is a power sum symmetric polynomial. In the sum (\ref{exp}) the term $m=1$ is excluded, and the partition $\beta$ has no parts equal to $1$.

Next, we write $p_\beta(X)=\sum_\rho \chi_\rho(\beta)s_\rho(X)$ and then join this Schur polynomial with the one already in the integrand, according to 
\be s_\rho(X)s_\lambda(X)=\sum_\nu c^\nu_{\lambda\rho}s_\nu,\ee
where $c^\nu_{\lambda,\rho}$ are the Littlewood-Richardson coefficients \cite{stanley}. The integral to be done is then
\be \frac{\mathcal{G}}{\mathcal{Z}}\int e^{-M(1-i\epsilon)\Tr(X)}|\Delta(X)|^2s_\nu(X)dX.\ee
But this is an integral of Selberg type \cite{selberg}, and is given by
\be \frac{d_\nu}{|\nu|!}([N]^\nu)^2[M(1-i\epsilon)]^{N^2-|\nu|}.\ee

The limit $N\to 0$ can be taken by noticing that, since $\lambda$ is a hook, we have \cite{novaessemi2} 
\be [N]^\lambda=Nt_\lambda+O(N^2).\ee This means only partitions $\nu$ that are also hooks will contribute and we get
\be C_n=\sum_{\lambda\in H_n}\frac{\chi_\lambda(n)}{t_\lambda^2}([M]^\lambda)^2B_\lambda,\ee where \be B_\lambda=\sum_{m} \sum_{\beta\vdash m}\frac{b_\beta(-M)^{\ell(\beta)}g_\beta(\epsilon)(n+m-1)!}{m!(n+m)[M(1-i\epsilon)]^{n+m}}D_{\lambda\beta},\ee
with \be D_{\lambda\beta}=\sum_{\rho\nu}\chi_\rho(\beta)\frac{c^\nu_{\lambda\rho}}{d_\nu}.\ee

For a given pair of hooks, $\lambda,\nu$, there are two different $\rho$ for which $c^\nu_{\lambda\rho}$ is not zero. If $\lambda=(n-k,1^k)$ and $\nu=(n+m-r,1^r)$, then $\rho_1=(m+k-r,1^{r-k})$ and $\rho_2=(m+k-r+1,1^{r-k-1})$. We thus have the sum
\be \sum_{\rho\in H_m}\chi_\rho(\beta)c^\nu_{\lambda\rho}=\chi_{\rho_1}(\beta)+\chi_{\rho_2}(\beta).\ee
It is a standard fact from representation theory that the restriction from $S_{n+1}$ to $S_n$ of $\chi_\lambda$ is the sum of $\chi_\rho$ over all partitions $\rho$ that result from the Young diagram of $\lambda$ by removing a box. Hence, the above sum equals $\chi_{\omega}(\beta,1)$ with $\omega=(m+k-r+1,1^{r-k})$. 

We now have to compute 
\be \sum_{\nu\in H_{n+m}}\frac{\chi_\omega(\beta,1)}{d_\nu}=\sum_{r=0}^{n+m-1}\chi_\omega(\beta,1)\frac{(n+m-r-1)!r!}{(n+m-1)!}.\ee Using that
\be \frac{(n+m-r-1)!r!}{(n+m)!}=\int_0^1 u^{r}(1-u)^{n+m-r-1}du\ee
we end up having to compute the sum 
\be \sum_{r=0}^{n+m-1}\chi_\omega(\beta,1)x^r,\ee
where $x=u/(1-u)$. But the characters $\chi_\omega$ have been studied \cite{Stanley} and it turns out that
\be \sum_{r=0}^{n+m-1}\chi_\omega(\beta,1)x^{\ell(\omega)}=xf_\beta(x),\ee
where
\be f_\beta(x)=\prod_{q \in \beta}[1-(-x)^q].\ee
Hence, 
\be \sum_{\nu}\frac{\chi_\omega(\beta,1)}{d_\nu(n+m)}=\int_0^1 u^k(1-u)^{n+m-k-1}f_\beta\left(\frac{u}{1-u}\right)du.\ee

Finally,
\be B_\lambda)=\sum_{m} \sum_{\beta\vdash m}\frac{b_\beta(-M)^{\ell(\beta)}g_\beta(\epsilon)(n+m-1)!}{m![M(1-i\epsilon)]^{n+m}}F_{n,m,k}(\beta)\ee
where
\be\label{F} F_{n,m,k}(\beta)=\int_0^1 u^k(1-u)^{n+m-k-1}f_\beta\left(\frac{u}{1-u}\right)du.\ee
Since $[M]^\lambda=(M-k)^{(n)}$ we get
\begin{widetext}
\be C_n=\sum_{k=0}^{n-1}\frac{(-1)^k}{t_\lambda^2}((M-k)^{(n)})^2\sum_{m} \sum_{\beta\vdash m}\frac{b_\beta(-M)^{\ell(\beta)}g_\beta(\epsilon)(n+m-1)!}{m![M(1-i\epsilon)]^{n+m}}F_{n,m,k}(\beta).\ee
\end{widetext}

This expression is very explicit and easy to implement in the computer. Even the integration in (\ref{F}) can be done exactly as it is always a Beta function. The first few orders in $1/M$ agree with the generating functions presented in \cite{berko2}.

\section{Correlator as power series in $\epsilon$}

Alternatively, we may express $C_n(M,\epsilon)$ as a power series in $\epsilon$. Such a series is not convergent: its radius of convergence cannot be finite because the integral in Eq.(\ref{I}) clearly does not exist if $\epsilon$ has a negative imaginary part. But the series can still be asymptotic and therefore useful, in the sense that its first $d$ terms give an accurate representation of the function for small $\epsilon$, up to an error of order $\epsilon^d$. 

After we expand
\be e^{Mi\epsilon \Tr(\frac{X}{1-X})}=\sum_{m=0}^\infty \frac{(iM\epsilon)^m}{m!}\sum_{\mu\vdash m}d_\mu s_\mu\left(\frac{X}{1-X}\right)\ee we arrive at
\be C_n=\lim_{N\to 0}\sum_{\lambda\in H_n}\chi_\lambda(n)\left(\frac{[M]^\lambda}{[N]^\lambda}\right)^2\sum_{m=0}^\infty \frac{(iM\epsilon)^m}{m!}\sum_{\mu\vdash m}d_\mu I_{\lambda\mu},\ee
where 
\be I_{\lambda\mu}=\frac{\mathcal{G}}{\mathcal{Z}}\int \det(1-X)^M|\Delta(X)|^2s_\mu\left(\frac{X}{1-X}\right)s_\lambda(X)dX.\ee 

Schur polynomials can be expressed as a ratio of determinants,
\be s_\lambda(X)=\frac{\det\left(x_j^{N+\lambda_i-i}\right)}{\det\left(x_j^{N-i}\right)}=\frac{\det\left(x_j^{N+\lambda_i-i}\right)}{\Delta(X)}.\ee
In the present case this gives
\be s_\lambda\left(\frac{X}{(1-X)}\right)=\det\left[\left(\frac{x_k}{(1- x_k)}\right)^{N+\lambda_i-i}\right]\frac{1}{\Delta\left(\frac{X}{(1- X)}\right)}.\ee
It is easy to express the above Vandermonde as
\be \Delta\left(\frac{X}{(1- X)}\right)=\frac{\Delta(X)}{\det(1- X)^{N-1}}.\ee
The integral can then be computed by means of the Andreief identity,
\be \int \det(f_i(x_k))\det(g_j(x_k))dX=N!\det\left[\int f_i(x)g_j(x)dx\right],\ee
which gives
\be I_{\lambda\mu}=N!\frac{\mathcal{G}}{\mathcal{Z}}\det\left[\int_0^1 (1-x)^{M-\mu_j+j-1}x^{2N+\mu_j-j+\lambda_i-i}dx\right],\ee
or
\begin{multline}\label{II} I_{\lambda\mu}=N!\frac{\mathcal{G}}{\mathcal{Z}}\prod_{j=1}^N\frac{(M-\mu_j+j-1)!}{(M+2N+\lambda_j-j)!}\\\times\det\left[(2N+\mu_j-j+\lambda_i-i)!\right].\end{multline}

The above determinant can be computed resorting again to the Andreief identity,
\begin{align} &\det\left[(2N+\mu_j-j+\lambda_i-i)!\right]\\&=\det\left[\int_0^\infty x^{2N+\mu_j-j+\lambda_i-i}e^{-x}dx\right]\\&=\frac{1}{N!}\int_0^\infty\det\left[x_j^{N+\lambda_i-i}\right]\det\left[x_i^{N+\mu_j-j}\right]e^{-\Tr(X)}dX \\&=\frac{1}{N!}\int_0^\infty s_\mu(X)s_\lambda(X)|\Delta(X)|^2e^{-\Tr(X)}dX\\&=\frac{1}{N!}\sum_\nu c_{\lambda\mu}^\nu\frac{d_\nu}{|\nu|!}([N]^\nu)^2.\end{align}

When we take $N\to 0$, we get that $\nu$ must be a hook, as well as $\mu$. We also recognize in (\ref{II}) an expression for the generalized falling factorial, so that 
\be C_n=\sum_{\lambda\in H_n}\frac{\chi_\lambda(n)}{t_\lambda^2}[M]^\lambda \widetilde{B}_\lambda,\ee with
\be \widetilde{B}_\lambda=\sum_{m=0}^\infty \frac{(iM\epsilon)^m}{m!}\sum_{\mu\in H_m} \frac{d_\mu}{[M]_\mu}\frac{(n+m-1)!}{(n+m)}\sum_{\nu} \frac{c^\nu_{\lambda\mu}}{d_\nu}.\ee
When $\lambda=(n-k,1^k)$ and $\mu=(m-l,1^l)$, the coefficient $c^\nu_{\lambda\mu}=1$ if and only if $\nu=(n+m-k-l,1^{k+l})$ or $\nu=(n+m-k-l-1,1^{k+l+1})$. Hence,
\begin{multline} \sum_{\nu\in H_{n+m}} \frac{c^\nu_{\lambda\mu}}{d_\nu}=\frac{(n+m-k-l-2)!(k+l)!(n+m)}{(n+m-1)!}\\=\frac{(n+m)}{(n+m-1)!}t_{\lambda\circ\mu},\end{multline}
where $\lambda\circ\mu=(n+m-k-l-1,1^{k+l}).$

Finally,
\be\label{CnM} C_n=\sum_{m=0}^\infty \frac{(iM\epsilon)^m}{m!}\sum_{\lambda\in H_{n}}\sum_{\mu\in H_{m}}\chi_\lambda(n)d_\mu\frac{t_{\lambda\circ\mu}}{t_\lambda^2}\frac{[M]^\lambda}{[M]_\mu}.\ee
This expression is of a different nature than the one obtained in the previous Section, but it is also very explicit and easy to implement.

\section{Statistics of time delay}

As discussed by Berkolaiko and Kuipers \cite{berko1}, the time delay moments
$Q_m=\langle\Tr(Q^m)\rangle$ can be obtained from appropriate derivatives of the energy correlators, 
\be\label{MfromC}
Q_m=\frac{M\tau_D^m}{i^mm!}\left[\frac{d^m}{d\epsilon^m}\sum_{n=1}^m
(-1)^{m-n}{m \choose n}C_n(\epsilon)\right]_{\epsilon=0}. \ee

Using the expression we have just derived for $C_n$ as a power series in $\epsilon$, Eq. (\ref{CnM}), it is easy to see that 
\be \frac{M}{i^m}\frac{d^mC_n}{d\epsilon^m}=M^m\sum_{\lambda\in H_n}\sum_{\mu\in H_m}\chi_\lambda(n) d_\mu \frac{[M]^\lambda}{[M]_\mu}\frac{t_{\lambda\circ\mu}}{t_\lambda^2 }.\ee From this we can write
\be Q_m=\frac{\tau_D^m}{i^mm!}\frac{d^m}{d\epsilon^m}\sum_{n=1}^m(-1)^{n+m}\binom{m}{n}C_n\ee
or
\be Q_m=(M\tau_D)^m\sum_{n=1}^m\binom{m}{n}\frac{(-1)^{n+m}}{m!}E_{nm},\ee
where
\be E_{nm}=\sum_{\lambda\in H_n}\sum_{\mu\in H_m}\chi_\lambda(n) d_\mu\frac{[M]^\lambda}{[M]_\mu}\frac{t_{\lambda\circ\mu}}{t_\lambda^2}.\ee

When $\lambda=(n-k,1^k)$ we have $[M]^\lambda=(M-k)^{(n)}$.
The sum over $n$ and the sum over $\lambda$ then becomes \cite{riedel}
\begin{widetext}
\be\label{reidel} \sum_{n=1}^m\sum_{k=0}^{n-1}\binom{m}{n}(-1)^{n-k}\frac{(n+m-k-l-2)!(k+l)!}{(n-k-1)!^2k!^2}(M-k)^{(n)}=(-1)^{m-l}(M-l)^{(m)},\ee\end{widetext}
with $\mu=(m-l,1^l)$. Therefore, we arrive at a very simple expression,
\be Q_m=\frac{(M\tau_D)^m}{m}\sum_{\mu \in H_m}\chi_\mu(m)d_\mu\frac{[M]^\mu}{[M]_\mu}.\ee
This coincides exactly, for any $n$, with the result derived from random matrix theory \cite{novaesrmt}.

\section{Conclusion}

Using a powerful semiclassical approach, based on matrix integrals, we investigated energy correlations in the scattering matrices of chaotic systems with broken time-reversal symmetry. We expressed the basic correlator $C_n(M,\epsilon)$, Eq.(\ref{corr}), in two different ways: as a power series in $1/M$ and as a power series in $\epsilon$. From the latter we were then able to extract average spectral moments of the time delay operator. We found complete agreement with RMT predictions, thereby microscopically justifying that approach. 

A natural extension of this work would be to perform analogous calculations for systems with intact time-reversal symmetry. That remains a challenge. Moreover, nonlinear statistics of time delay, like $\langle [\Tr(Q)]^n\rangle$, have been computed within RMT, but are not accessible to the present approach. We believe the alternative semiclassical treatment introduced in \cite{KSS} is promising in that respect.

\section*{Acknowledgments}
Financial support from CNPq, grant 306765/2018-7, is gratefully acknowledged. I would like to thank Marko Riedel for providing a proof of equation (\ref{reidel}).

\end{document}